\documentclass[aps,prl,reprint,twocolumn,superscriptaddress,amsmath,notitlepage]{revtex4-1}

\usepackage{graphicx,bm,wasysym,color}
\usepackage{amsmath,amsfonts,amsthm,amssymb,amsbsy,bbold,mathrsfs}
\usepackage[caption=false]{subfig}

\newcommand{\nonarxivonly}[1]{}
\def \li213 {Li$_2$IrO$_3$ }
\def \na213 {Na$_2$IrO$_3$ }

\begin{document}
%%%%%   TITLE PAGE   %%%%%%
\title{Unified theory of spiral magnetism in the harmonic-honeycomb iridates
$\alpha$, $\beta$, $\gamma$ Li$_2$IrO$_3$}
\author{Itamar Kimchi}
\affiliation{Department of Physics, University of California, Berkeley, CA
94720, USA}
\author{Radu Coldea}
\affiliation{Clarendon Laboratory, University of Oxford, Parks Road, Oxford
OX1 3PU, U.K.}
\author{Ashvin Vishwanath}
\affiliation{Department of Physics, University of California, Berkeley, CA
94720, USA}
\affiliation{Materials Science Division, Lawrence Berkeley National
Laboratories, Berkeley, CA 94720, USA}
\begin{abstract}
A family of insulating iridates with chemical formula
Li$_2$IrO$_3$ has recently been discovered, featuring three
distinct crystal structures $\alpha,\beta,\gamma$ (honeycomb,
hyperhoneycomb, stripyhoneycomb). Measurements on the
three-dimensional polytypes, $\beta$- and $\gamma$-Li$_2$IrO$_3$,
found that they magnetically order into remarkably similar spiral
phases, exhibiting a non-coplanar counter-rotating spiral magnetic
order with equivalent $q=0.57$ wavevectors. We examine magnetic
Hamiltonians for this family and show that the same triplet of
nearest-neighbor Kitaev-Heisenberg-Ising ($KJI$) interactions
reproduces this spiral order on both $\beta,\gamma$-Li$_2$IrO$_3$
structures. We analyze the origin of this phenomenon by studying
the model on a 1D zigzag chain, a structural unit common to the
three polytypes. The zigzag-chain solution transparently shows how
the Kitaev interaction stabilizes the counter-rotating spiral,
which is shown to persist on restoring the inter-chain coupling.
Our minimal model makes a concrete prediction for the magnetic
order in $\alpha$-Li$_2$IrO$_3$.
\end{abstract}

%%%%%   MAIN TEXT   %%%%%%
\maketitle

%%%%%   FIGURES   %%%%%%
\begin{figure}[b]
\includegraphics[width=0.95\columnwidth]{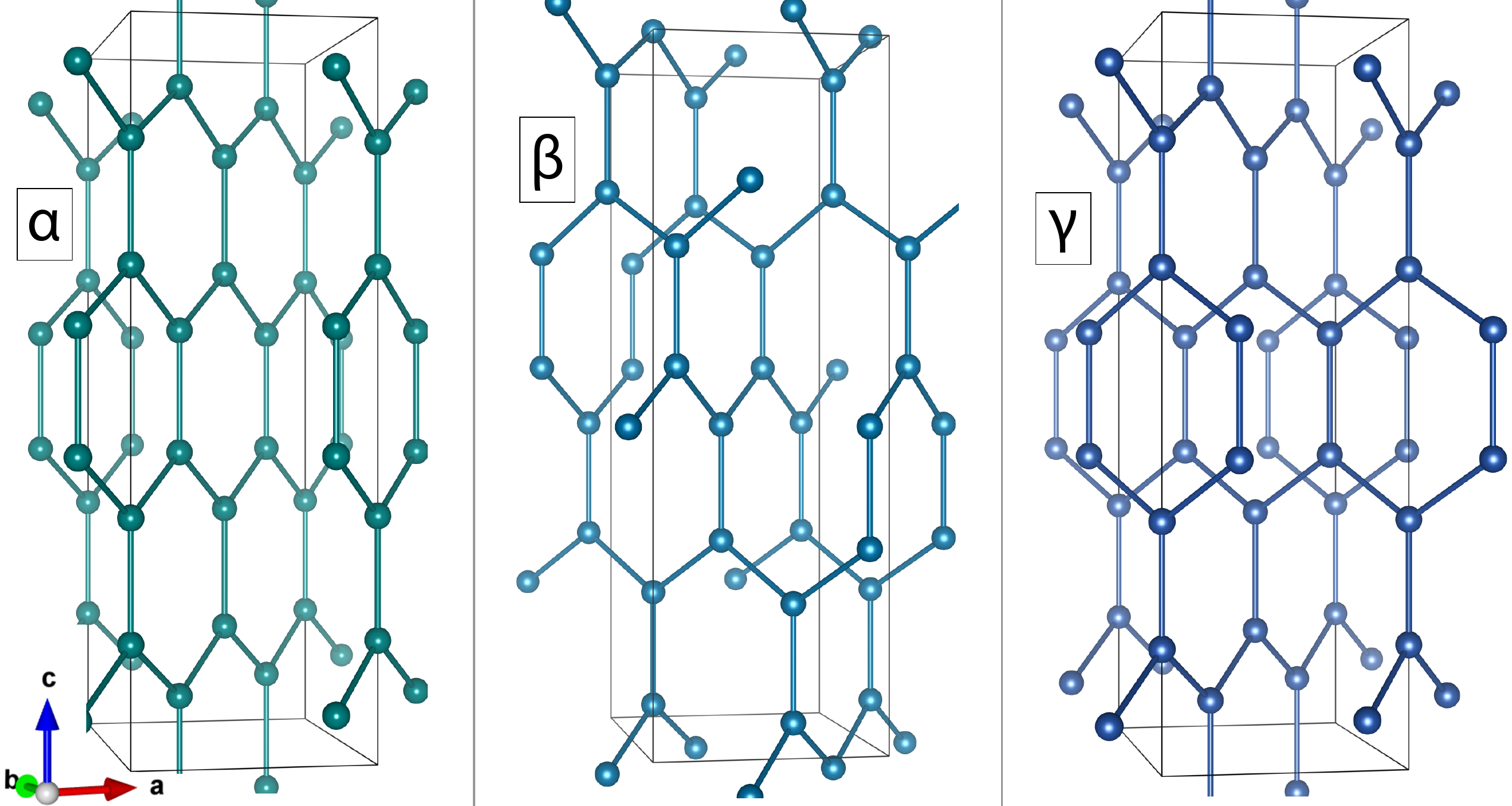}
\caption[]{{\bf Lattices} of Ir in
$\alpha,\beta,\gamma$-Li$_2$IrO$_3$, with parent orthorhombic
$a,b,c$ axes. Experiments on the 3D lattices, $\beta$- and
$\gamma$-Li$_2$IrO$_3$, found strikingly similar spiral orders. }
\label{fig:lattices}
\end{figure}

\begin{figure*}[ht]
\includegraphics[width=\textwidth]{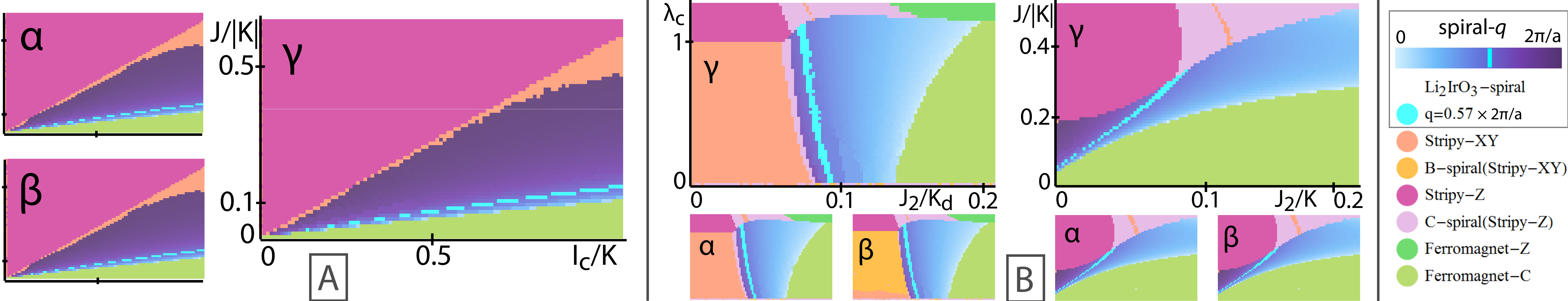}
\caption[]{{\bf Phase diagrams on $\alpha,\beta,\gamma$-Li$_2$IrO$_3$.} In the
vicinity of the spiral phase (shaded blue) which contains the experimentally
observed magnetic order, the semiclassical phase diagrams appear remarkably
similar across the $\alpha,\beta,\gamma$-Li$_2$IrO$_3$ lattices. \\
(A) The nearest-neighbor $KJI_c$ model ($J_2{=}0$) is sufficient for capturing
the observed spiral, and exhibits this cross-lattice similarity.
(B) (Left) the spiral from the 1D zigzag chain model persists to the full
lattices; (right) taking $J_2 \rightarrow 0$ requires large $|K|/J$; see
parameters below.
For the 2D $\alpha$-polytype, shading indicates the equivalent spiral $\bm{q}$
along $\bm{a}$ as described in the text.
}
\label{fig:phases}
\end{figure*}

\begin{figure}[]
\includegraphics[width=\columnwidth]{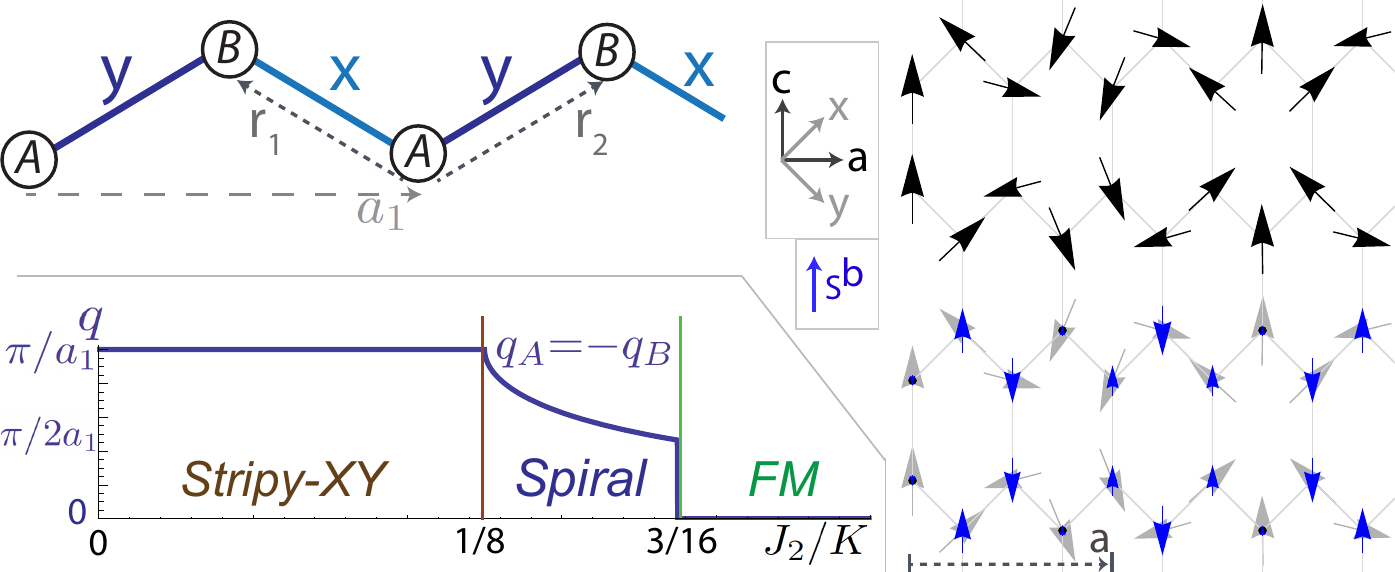}
\caption[]{{\bf Zigzag chain and spiral.}  As evident in this 1D
minimal model for the Li$_2$IrO$_3$ lattices (top left), the
counter-rotating coplanar spiral order can be stabilized by Kitaev
interactions within the coplanar ansatz Eq.~\ref{eq:spinansatz} (bottom left; here with  $K{<}0$, $J{=}|K|/3$). For
each lattice, restoring the inter-chain couplings preserves the
counter-rotating $S^a,S^c$ spiral (top right), while also
introducing non-coplanar  $S^b$ components (overlayed in blue,
bottom right).  Together they form the experimentally observed
order. } \label{fig:zigzag_phases}
\end{figure}

%%%%%   INTRODUCTION   %%%%%%
Edge-sharing oxygen octahedra coordinating Ir$^{4+}$ ions can
exhibit unconventional magnetic interactions between the Ir
$S_{\text{eff}}{=}1/2$ pseudospins. Strong spin orbit coupling in
iridium, which produces these low energy Kramer's doublets, can
combine with 90$^\circ$ Ir-O-Ir exchange pathways to generate
bond-dependent couplings identical to those discussed by
Kitaev\cite{Kitaev2006}, as has been proposed in
Refs.~\onlinecite{Khaliullin2009} and \onlinecite{Khaliullin2010}
for Na$_2$IrO$_3$. The collinear antiferromagnetic
magnetism\cite{Gegenwart2010, Hill2011,Cao2012,Taylor2012} later
found in Na$_2$IrO$_3$ is distinct from simple Neel order, but can
be captured by various models with or without Kitaev-type spin
anisotropies.\cite{Gegenwart2012,Taylor2012,You2011,Lauchli2011,Khaliullin2012,Damascelli2012,Valenti2013,Valenti2013a,MazinPrivate,Min2013,Kim2013a,Kim2013,Daghofer2012comment}
The isostructural compound $\alpha$-Li$_2$IrO$_3$, in which Ir
forms separated layers of the 2D honeycomb lattice, is available
only in powder form. Thermodynamic and susceptibility measurements
suggest it also orders magnetically\cite{Gegenwart2012}, and
powder neutron diffraction experiments found a magnetic Bragg peak
with a small nonzero wavevector inside the first Brillouin
zone\cite{RaduDresden}, stimulating theoretical
models\cite{Rachel2014,vandenBrink2014a} of spiral orders.

In the past two years, compounds with chemical formula
Li$_2$IrO$_3$ have been successfully synthesized in two additional
crystal structures (Fig.~\ref{fig:lattices}). In
$\gamma$-Li$_2$IrO$_3$ the Ir sites form the 3D stripyhoneycomb
lattice\cite{Analytis2014,Coldea2014} (space group \#66 $Cccm$),
featuring hexagons which are arranged in honeycomb strips of
alternating orientation. In $\beta$-Li$_2$IrO$_3$ the Ir sites
form the 3D hyperhoneycomb lattice\cite{Takagi2014,Coldea2014a}
(space group \#70 $Fddd$), featuring 10-site decagons which are
reminiscent of the hyperkagome\cite{Takagi2007} lattice of
Na$_4$Ir$_3$O$_8$. The relation between these structures is
captured by their designation as harmonic-honeycomb
iridates\cite{Analytis2014,Kitaev3D}, a structural series in which
$\alpha,\beta,\gamma$-Li$_2$IrO$_3$ are labelled by $n=\infty,0,1$
respectively.  Common features include local three-fold
coordination of sites, as well as identical 2D projections along
the $\bm{a}$ and $\bm{b}$ parent orthorhombic axes; the $\bm{c}$
axis projections are distinct.

Recent experiments using resonant magnetic x-ray diffraction have
successfully determined the magnetic ordering in $\beta$- and
$\gamma$-Li$_2$IrO$_3$ single
crystals\cite{Coldea2014,Coldea2014a}.  The results are striking.
Both compounds order into a  complex spiral at a temperature
$T_N{=}$38 K.  This order hosts counter-rotating spirals within
the unit cell, exhibiting a particular pattern of non-coplanar
tilts. The spiral wavevector $\bm{q}$ lies along the orthorhombic
$\bm{a}$ axis, with the same apparently incommensurate magnitude
$q =0.57(1){\times} 2\pi/a = 0.61(1)$\AA$^{-1}$ in both
structures. Except for the angle of the non-coplanar tilt,  the
magnetic orders observed in  $\beta$- and $\gamma$-Li$_2$IrO$_3$
are equivalent to each other, though occuring in different lattice
settings.

In this work we analyze the origin of this phenomenon by
theoretically studying the three Li$_2$IrO$_3$ systems at the
level of lattice magnetic Hamiltonians. We show that a
microscopically-derivable set of nearest-neighbor interactions,
consisting of Kitaev, Heisenberg and Ising exchanges, is
sufficient for capturing the observed spiral magnetic order. This
Hamiltonian is
\begin{align}
\label{eq:KJI}
H& = \sum_{\langle i j \rangle} \left[K\ S_i^{\gamma_{i j}} S_j^{\gamma_{i j}}
+ J\ \vec{S}_i\cdot \vec{S}_j + I_c\ S_i^{r_{i j}} S_j^{r_{i j}} \right]
\end{align}
where $K$ is the Kitaev coupling, and $I$ is a distinct Ising
coupling of the spin components parallel to the bond orientation,
i.e. $S^{r_{i j}} {\equiv} \vec{S} {\cdot} \hat{r}_{i j}$ where
$\hat{r}_{i j}{=} (\vec{i} {-} \vec{j})/| i {-}j|$ is the unit
vector from site $i$ to site $j$ (see the Appendix including
Fig.~\ref{fig:octahedra} for details). In this model the Ising term
$I_c$ is chosen to be active only on those symmetry-distinguished
bonds which are parallel to the $\bm{c}$ axis, where it becomes
$I_c S^c_i S^c_j$. For the Kitaev coupling of spin component
$\gamma_{i j}$, the bond-dependent axis $\gamma_{i j}\in(x,y,z)$
is the Ir-O unit vector from iridium site $i$ to one of the
oxygens in its coordinating octahedron, uniquely chosen so that
$\gamma_{i j}$ is perpendicular to $r_{i j}$ or, equivalently,
perpendicular to the bond's Ir$_i$O$_2$Ir$_j$ square. Here
$\hat{z}{=}\hat{b}$ and
$\hat{x},\hat{y}{=}(\hat{a}{\pm}\hat{c})/\sqrt{2}$. As is clear
from this representation, the three different exchanges $K,J,I$
are all symmetry-allowed and can be microscopically
generated\footnote{Microscopic exchange pathways for edge-sharing
octahedra have been discussed in Refs.
\onlinecite{Khaliullin2005,Balents2008,Khaliullin2009,Khaliullin2010,Norman2010,Micklitz2010,Khaliullin2012,Khomskii2012,Valenti2013,Kee2014}.}
already in the limit of cubic O$_6$ octahedra.

The phase diagram of Eq.~\ref{eq:KJI}, shown in
Fig.~\ref{fig:phases}A, exhibits a remarkable feature.  The
experimentally-observed spiral order in the $\beta$ and $\gamma$
lattices is stabilized in our theoretical model as the ground
state on all three lattices, for certain parameters such as
$(K,J,I_c) = (-12,0.6,-4.5)$ meV. Moreover the surrounding phase
diagrams, computed (see details below) by setting Eq.~\ref{eq:KJI} on each of
the
three $\alpha,\beta,\gamma$-Li$_2$IrO$_3$ lattices, are all quite
similar. In Fig.~\ref{fig:phases} the phase diagrams on
$\alpha,\beta,\gamma$ lattices are shown for the same parameter
range, permitting this visual comparison. This feature suggests
that the experimental observations, of the striking similarity
between the $\beta$- and $\gamma$-Li$_2$IrO$_3$ spiral orders, may
be captured within this effective  $S{=}1/2$ Hamiltonian with
nearest-neighbor exchanges.

To understand the striking similarity between the
Fig.~\ref{fig:phases} phase diagrams found in our numerical
computations on the different lattices, we introduce a conceptual
toy model consisting of a 1D zigzag chain. This minimal conceptual
model may be motivated as follows. Observe that the symmetries of
the Li$_2$IrO$_3$ polytypes single out the set of Ir-Ir bonds
which lie parallel to the crystallographic $\bm{c}$ axis. These
$c$-bonds, with $r_{i j}{=}c$, all carry Kitaev couplings of
$\gamma_{i j} {=} z{=}b$. The remaining ``$d$-bonds'' (as well as
their $\gamma_{i j} {=} x,y$) all lie diagonal to the $a,b,c$
axes. This symmetry-enforced distinction suggests the microscopic
mechanisms for setting $I_d{=}0$ in Eq.~\ref{eq:KJI}. Now consider
decomposing the Hamiltonian Eq.~\ref{eq:KJI} into its interactions
on $c$-bonds and on $d$-bonds, $H=H_c+H_d$.  The $d$-bonds
Hamiltonian $H_d$ is then a sum of decoupled 1D zigzag chains at
various positions and orientations, $H_d=\sum H_{\text{1D}}$,
turning all three lattices into sums over identical $
H_{\text{1D}}$  building blocks.

%%%%%   Zigzag chain   %%%%%%
\textbf{Zigzag chain minimal model.} The zigzag chain toy model is
a conceptual mechanism for connecting the full numerical
computations. Its solution is transparent, clarifying how
essentially the same form of spiral order arises from
Eq.~\ref{eq:KJI} on the distinct 3D lattices. We complement its
analytical insight by numerically computing the phase diagrams as
we mathematically interpolate between the 3D lattices: even as we
smoothly turn off the inter-chain bonds, reducing the 3D lattices
to the 1D chain, the spiral phase remains stable.

Since we define $H_{\text{1D}}$ by dropping the inter-chain
$c$-bonds, we here mitigate the loss of the $I_c$ exchange by
introducing a second-neighbor Heisenberg $J_2$ interaction. This
$J_2$ can be discarded when the full 3D lattice is restored. The
zigzag-chain geometry is defined in Fig.~\ref{fig:zigzag_phases};
let $r_1,r_2$ point from an $A$-sublattice site to its neighboring
$B$ sites, and choose the 1D Bravais lattice with vector
$a_1=r_2{-}r_1$ so that the $A$-sites lie at integer positions
$r=n a_1$. The single-chain Hamiltonian is
\begin{align}
\label{eq:chain}
H_{\text{1D}} = \sum_{r{=}n a_1} & \bigg[\
\, K\ \, \big(S_{A,r}^x \ S_{B,r{+}r_1}^x + S_{A,r}^y\ S_{B,r{+}r_2}^y \big)
\\
&+ J\ \, \big(\vec{S}_{A,r}\cdot \vec{S}_{B,r{+}r_1} + \vec{S}_{A,r} \cdot
\vec{S}_{B,r{+}r_2} \big)\phantom{\big[} \notag\\
& + J_2 \, \big(\vec{S}_{A,r}\cdot \vec{S}_{A,r{+}a_1} +
\vec{S}_{B,r{-}r_2}{\cdot} \vec{S}_{B,r{+}r_1} \big) \bigg] \notag
\end{align}

In the following we consider the $x,y$ coplanar spiral magnetic orders that could be stabilized by the 1D minimal model Eq.~\ref{eq:chain}, with spin ordering confined to the $x,y$ (or equivalently $a,c$) plane. (Restoring the
inter-chain  $z$-type Kitaev couplings will produce the
non-coplanar tilt.)  
%This is justified perturbatively near an exactly solvable point, as follows.
First consider Eq.~\ref{eq:chain} at the exactly solvable point  $J_2{=}0$,
 $K{=}{-}2J$,
$K{<}0$, where a site-dependent spin
rotation\cite{Khaliullin2005,Khaliullin2010,KHbhc} exposes it as a
Heisenberg ferromagnet in a rotated basis. Its exact quantum ground
state is a \textit{Stripy} collinear antiferromagnet (AFM) of the
original spins. 
%In particular, taking now slightly larger $|K|$, the ground state is Stripy-XY: it has spins collinear along
Now perturbing around this point by taking $J$ smaller than $|K|/2$, the ground state has Stripy-XY antiferromagnetic order, with ordered spins collinear along
$S^x/S^y$ which are aligned on $x/y$-type bonds and anti-aligned
along  $y/x$-type bonds. Focusing on large FM $K{<}0$ with small
AF $J{>} 0$ satisfying $K{+}2J {<}0$, we expect the zigzag chain 
model to capture states which are $x,y$-coplanar.

%We therefore conceptually consider an ansatz for the $x,y$ spin components, on the sublattices $s\in\{A,B\}$,
Switching on $J_2$ frustrates the stripy order. Focusing on the scenarios where the ordered spins remain confined to the $x,y$ plane, we conceptually consider below a general spiral order with the ansatz 
\begin{equation}
\label{eq:spinansatz}
\left( S_{s,r}^x, S_{s,r}^y \right) =
\big(\text{Re},\text{Im}\big)\left[\exp\{-i\left( q_s\, r+\phi_s \right)\}
\right]
\end{equation}
where the sublattice index $s$ takes the values $\{A,B\}$, and the two sublattices have spiral wavevectors $q_A{=}{\pm}q_B$ and  phases $\phi_A,\phi_B$.

Consider the  case of counter-rotation,  $q_B{=}-q_A{=}\theta/a_1$
with $\theta{ >} 0$ ($a_1$ is defined in Fig.~\ref{fig:zigzag_phases}).
The energy per unit cell is given by
\begin{equation}
\label{eq:Eminus}
E_{-}(\theta)=K\sin\left(\theta/2\right)\sin(\phi_A{+}\phi_B)+2J_2\cos(\theta)\end{equation}
Minimizing the energy with respect to the sublattice phases (for
$K{<}0$) immediately fixes their sum to be
$\phi_A{+}\phi_B=\pi/2$. Now consider the minimization with
respect to the spiral rotation angle $\theta$. There are three
cases.  (1) For small $|J_2|$, Eq.~\ref{eq:Eminus} is minimized at
$\theta=\pi$, producing the Stripy-XY AFM state, with energy
$E_{\text{stripy}}=K{-}2J_2$. (2) For larger ferromagnetic
$J_2<0$, a global minimum develops at an incommensurate wavevector
fixed by $\sin\left(\theta/2\right) {=} K/(8J_2)$, for
$|J_2|>|K|/8$. This incommensurate counter-rotating spiral phase
has energy $E_{\text{spiral}}=2J_2 + K^2/(16J_2)$. (3) At larger
$|J_2|$ it gives way to the $q{=}0$ ferromagnet  solution
($\phi_A{=}\phi_B$) with energy $E_{FM}=K{+}2J{+}2J_2$. The 
 phase diagram as well as the associated wavevector $q$, which result from this computation with the coplanar ansatz Eq.~\ref{eq:spinansatz},   are shown in
Fig.~\ref{fig:zigzag_phases}.

It is also evident that a mostly-Heisenberg model cannot produce a
counter-rotating spiral.  This is true even if it is supplemented
by e.g.\ Dzyaloshinskii-Moriya  couplings. To see this, examine
the generic spin correlations of the ansatz state
Eq.~\ref{eq:spinansatz}. Between neighboring sites $i{=}(A,r)$ and
$j{=}(B,r{+}v)$, they are
\begin{align}
\label{eq:spincorrelations}
\left\langle S_i^x S_j^x \pm S_i^y S_j^y \right\rangle = \delta(q_B \mp q_A)
\cos\left( q_B v + \phi_B \mp \phi_A \right)
\end{align}
The upper sign gives the usual Heisenberg correlations, while the
lower sign corresponds to the spin-anisotropic correlations of the
Kitaev exchange. The delta-function factor ensures that the
Heisenberg/Kitaev correlations vanish in the counter/co-rotating
spiral, respectively.

\textbf{Non-coplanar spiral from coupled chains.} Each of the
three  $\alpha,\beta,\gamma$-Li$_2$IrO$_3$ lattices is reached
from the decoupled-chains limit, by introducing a particular
pattern of inter-chain couplings between chains of various
positions and orientations.  We find that these inter-chain
couplings both help to stabilize the coplanar spiral found in the
1D model, and also introduce an alternating pattern of
non-coplanar tilts in the rotation planes of successive zigzag
chains, as follows. By taking Eq. \ref{eq:spinansatz} with
appropriate phases and introducing the  $\langle S^b \rangle $
component, we describe the full spiral by
\begin{equation}
\label{eq:spinfn}
\vec{S}_{s,r}=\cos(q_s r_a)\langle S^c \rangle \hat{c}- \sin(q_s
r_a)\left(\langle S^a \rangle\hat{a} \pm \langle S^b \rangle\hat{b} \right)
\end{equation}
with $q_B{=}{-}q_A{=}q{>}0$ denoting counter-rotation between
upper ($s{=}B$) and lower ($A$) sites on each zigzag chain. The
$\pm$ sign alternates between successive zigzag chains, tilting
$S^a$ towards ${\pm} S^b$, with magnitudes satisfying $\langle S^a
\rangle^2{+}\langle S^b \rangle^2{=}\langle S^c \rangle^2$
required by the constraint of fixed length spin on each site. This
tilting is stabilized energetically by the strong $K_c S^b_i
S^b_j$ inter-chain coupling, and its alternating pattern is set by
$J_c{>}0$.  The resulting non-coplanar spiral is composed of a
coplanar spiral in each zigzag chain, whose plane of rotation
alternates in orientation between adjacent zigzag chains.
Fig.~\ref{fig:zigzag_phases} shows the resulting  spiral as viewed
in the $b$-axis projection common to the lattices, for parameters
with  $q=0.57{\times}2\pi/a$.

\textbf{Applicability of the 1D model.} We demonstrate the
applicability of the 1D model to the physical lattices, by
studying the smooth evolution of each lattice to its
decoupled-chains limit.  In particular, we introduce an
inter-chain coupling coefficient $\lambda_c$, and map the
semiclassical phase diagram of $H_\lambda=\lambda_c H_c+H_d$.
Here the Hamiltonian Eq.~\ref{eq:KJI} is supplemented by the $J_2$
exchange between second-neighbors of the Ir lattice, on the two
intra-chain bonds (as in Eq.~\ref{eq:chain}) as well as on the
four remaining bonds (where it is suppressed by the inter-chain
coupling coefficient $\lambda_c$). Such a study is shown in
Fig.~\ref{fig:phases}B, showing the phase diagram as a function of
$\lambda_c$ and $J_2$ for $K_d{=}0.8K_c$,
$J_c{=}2J_d{=}|I_c|,I_c{=}K_c{/}3$. These parameters, though not likely to be
physically relevant, allow this mathematical interpolation from 3D to 1D. We
find that the spiral phase
remains stable from the 1D limit $\lambda_c{=}0$ through the
isotropic physical lattice $\lambda_c{=}1$, on each of the
lattices.

\textbf{Necessity of strong Kitaev interactions.} We consider a
$KJI_c{-}J_2$ Hamiltonian, such as the model we previously
reported\cite{Coldea2014} for the spiral order in
$\gamma$-Li$_2$IrO$_3$, and attempt to tune $J_2\rightarrow 0$
while preserving the experimentally-observed spiral phase.  Such a
study is presented in  Fig.~\ref{fig:phases}B, showing the phase
diagram in $J/|K|$ and $J_2/K$, here for $I_c/K=0.375$.  We find
that to discard the second neighbor interactions, the ratio
$|K|/J$ must simultaneously be taken to be quite large $\sim 20$.
One representative such set of nearest-neighbor couplings is
$(K,J,I_c) = (-12,0.6,-4.5)$ meV. Here the overall scale is set so
that the mean field ordering temperature $T_N{=}40$K matches the
experimental $T_N$. Putting aside the Ising term, this ratio
$J/|K|=0.05$ lies well within the 2D Kitaev quantum spin liquid
phase on the honeycomb
lattice\cite{Khaliullin2010,Khaliullin2012,Trebst2011b}, though it
may lie outside the 3D quantum spin liquid phases on the 3D
lattices\cite{Kitaev3D}.

\textbf{Semiclassical solutions.} The semiclassical approximation
which we employ can capture  incommensurate spiral orders as well
as other magnetic phases. 
We represent spins by unconstrained vectors, yielding a quadratic Hamiltonian
which is appropriate for capturing fluctuating states with small ordered
moments. The lowest energy mode of this quadratic Hamiltonian is associated
with the ordering instability of the spin model, and is straightforwardly
found by Fourier transform.
This is expected to be the leading ordering instability out of a high
temperature paramagnetic phase assuming a continuous transition. Potentially
quantum fluctuations could play a similar role. Our phase diagrams outline the
evolution of this leading instability.

The algorithmically-generated phase diagrams in
Fig.~\ref{fig:phases} host the Li$_2$IrO$_3$ spiral phase as well
as various competing orders. These include stripy
antiferromagnets, where spins of the given component are aligned
only along that Kitaev bond type; incommensurate orders with
$q$-vectors along $\bm{b}$ or $\bm{c}$, which retain stripy-like
correlations within the unit cell; and ferromagnets with $S^c$ or
$S^z$ alignment.

\textbf{Coplanar and tilt modes.} The experimentally observed
spiral phase in the $\beta$ and $\gamma$ lattices, expressed in
Eq.~\ref{eq:spinfn} and plotted in Fig.~\ref{fig:zigzag_phases},
was identified numerically in two steps. Observe that the
non-coplanar $S^b$ tilt pattern is distinguished from the $ S^a ,
S^c $ coplanar spiral order by a mirror eigenvalue, associated
with a $\bm{c}$-axis reflection. The coplanar spiral is
mirror-even while the tilt mode is mirror-odd. Indeed we find that
they appear as distinct modes in the Fourier transform of
Hamiltonians in the spiral phase. The global ground state is numerically found to be the coplanar spiral mode, which furthermore is found to
exhibit $\langle S^a \rangle < \langle S^c \rangle$. Nonlinear effects above
our quadratic approximation, which would tend to force the length of spin to
be similar across sites, are likely to mix this solution with an additional
mode.
We adopt the following heuristic approach to include effects of nonlinearity
which become more important with growing magnitude of the order parameter. We
examine the lowest energy excited mode available for this mixing, and find
throughout that it consists of the experimentally-observed $\langle S^b
\rangle$ tilt
pattern. While the instability analysis provides us a phase diagram that
includes an incommensurate spiral, a more controlled calculation of nonlinear
effects is required to decide whether the observed magnetic order appears or
some other state is favored in this regime of parameters for the quantum S=1/2
Hamiltonian.

This analysis fixes the pattern of non-coplanar
tilts. Their rough magnitude (though not their overall sign) can be estimated
by constructing a fully-classical configuration from the two mixing modes. For
the values $(K,J,I_c)
{=} ({-}12,0.6,{-}4.5)$ meV, the resulting tilt angle is
$63^\circ$, similar to the angles observed experimentally,
$42^\circ$ and $55^\circ$; it can be tuned through these values by
varying the relative ratios of the exchange parameters. However we
expect fluctuations to be relevant for these systems. Indeed,
in the experimentally-determined magnetic
structures\cite{Coldea2014,Coldea2014a} of $\beta$- and
$\gamma$-Li$_2$IrO$_3$,
the extracted ordered magnetic moment is not constant in magnitude
between sites, but it is smaller by 10-20\% when it is aligned in
the $ab$ plane compared to when it is pointing along the $c$-axis.
This variation is likely due to a combination of g-factor
anisotropies and quantum fluctuations of these $S{=}1/2$ moments.

% Ising interactions

\textbf{Zigzag-chain mechanism in $\alpha$-Li$_2$IrO$_3$.}
$\alpha$-Li$_2$IrO$_3$ \cite{OMalley2008} has a layered structure
of stacked 2D iridium honeycombs separated by layers of Li ions.
For comparison with the other lattices we construct an
orthorhombic parent unit cell of the same size as for the $\beta$
and $\gamma$ structures (see the Appendix for details) where the
honeycombs are in the ($\bm{a}{+}\bm{b}$,$\bm{c}$) plane
(Fig.~\ref{fig:lattices}). The Hamiltonian Eq.~\ref{eq:KJI} predicts an
incommensurate spiral order in the honeycomb layers with the same pattern of
counter-rotation between adjacent sites and non-coplanarity between vertical
($c$-axis) bonds as in the $\beta$ and $\gamma$ lattices. Remarkably, the
energetics is such that for the same values of the exchange
parameters ($K,J,I$), the calculated relative angles of spins on
nearest-neighbor sites is the same on all three lattices.

In particular, energetic analysis of the $(K,J,I)$ model Hamiltonian on the
$\alpha$-Li$_2$IrO$_3$ lattice, with parameters chosen to reproduce the
experimentally-observed order on $\beta$- and $\gamma$-Li$_2$IrO$_3$, predicts
a magnetic structure where the relative spin orientations between adjacent
sites are the same as in the $\beta$ and $\gamma$ polytypes.
This implies that the projection of the $\alpha$-Li$_2$IrO$_3$ ordering
wavevector onto the honeycomb layers is
$q_{\text{1D}} = q \cos \theta$, where $q=0.57*2\pi/a$ is the
propagation vector magnitude in the $\beta$ and
$\gamma$ lattices, and $\theta=\cos^{-1}(a/\sqrt{a^2+b^2})$ is the
angle between the $\bm{a}$-axis and the $\alpha$-Li$_2$IrO$_3$ honeycomb
layers.
Here the subscript $\text{1D}$ emphasizes that for a given honeycomb plane,
the spiral wavevector lies along a zigzag chain, as in the 1D model of
decoupled chains (Eq.~\ref{eq:chain} and Fig.~\ref{fig:zigzag_phases}).

The resulting value for this projection, $q_{\text{1D}}~\sim 0.35$\AA$^{-1}$,
serves as an estimated lower bound for the magnitude of the 3D ordering
wavevector ${\bm q}_{\rm{3D}}$ that would occur in the real material. Weak
inter-layer couplings can give ${\bm q}_{\rm{3D}}$ a finite component normal
to the honeycomb layers, suggesting a possible range for the magnitude $|{\bm
q}_{\rm{3D}}|$.
Future experiments on $\alpha$-Li$_2$IrO$_3$ single crystal samples could test
these predictions for ${\bm q}_{\rm{3D}}$, as well as the predictions for
non-coplanarity and counter-rotation, which are highly non-trivial features
for the magnetic order on a honeycomb lattice. In particular the
non-coplanarity would break the $C$-centering of the honeycomb
lattice, leading to a doubling of the primitive unit cell; this is a rather
unusual feature for spiral order, and distinct from other
theoretical models\cite{Rachel2014,vandenBrink2014a} for
$\alpha$-Li$_2$IrO$_3$.

\textbf{Conclusion.} The experimental observations in $\beta$- and
$\gamma$-Li$_2$IrO$_3$ are intriguing: the two compounds undergo a
magnetic ordering transition, at similar temperatures, into an
unusual spiral magnetic order, with spiral wavevectors which are
the same up to the experimental accuracy. This spiral wavevector
appears to be incommensurate, with no clear mechanism for strong
lattice pinning. In this work we have found a nearest-neighbor
magnetic Hamiltonian which reproduces the complete symmetry of the
spiral magnetic order on the two lattices including the pattern of
counterrotation and noncoplanarity. The origin of this
cross-lattice similarity is clarified by a 1D zigzag chain minimal
model. This transparent model is sufficiently minimal to be a
common building-block for the lattices, yet sufficiently complex
to stabilize the counter-rotating spiral order. Its applicability
is verified by smoothly extending it towards the physical
lattices, and its predictions for $\alpha$-Li$_2$IrO$_3$ are
testable. The apparent commonality across the Li$_2$IrO$_3$ family
suggests that to capture certain aspects of the magnetism, it may
be sufficient to describe the different compounds via the same
low-energy effective Hamiltonian. Why this may happen remains to
be understood.

\textbf{Note added.} During publication of this manuscript, a
preprint\cite{Kim2014} has appeared which discusses magnetism on the
$\beta,\gamma$-Li$_2$IrO$_3$ lattices.
One of the magnetic spiral phases identified there correctly captures the
magnetic structure observed\cite{Coldea2014a} in $\beta$-Li$_2$IrO$_3$.
However, that phase, as well as the other spiral phases found in that work,
differ in detail (symmetry of the
ordering pattern)
\footnote{
The $a$-axis spiral orders discussed in Ref.\ \onlinecite{Kim2014}
(``SP$_{a^+}$'' and ``SP$_{a^-}$'') exhibit features of non-coplanarity and
counter-rotation, but have a different symmetry compared to the spiral phase
found experimentally in $\gamma$-Li$_2$IrO$_3$. The pattern of non-coplanarity
of the spiral planes predicted for $\gamma$-Li$_2$IrO$_3$ is such that it
alternates between successive {\em pairs} of zigzag chains along $c$; whereas
experimentally it is found that it alternates between consecutive zigzag
chains\cite{Coldea2014}, as illustrated in Fig.~3 (bottom right). The order in
$\beta$-Li$_2$IrO$_3$ is however correctly captured by one of the spiral
orders found in that work with sign-flipped $\Gamma$-interactions,
specifically $\bar{SP}_{a^-}$. In contrast, here the experimentally-determined
structures for both the $\beta$ and the $\gamma$ polytypes are captured
naturally by the $KJI$ minimal model proposed in Eq.~1.}
from the spiral phase discussed here and observed
experimentally\cite{Coldea2014} for $\gamma$-Li$_2$IrO$_3$.

\textbf{Acknowledgments.} We thank James Analytis for previous
collaborations. This work was supported by the Director, Office of
Science, Office of Basic Energy Sciences, Materials Sciences and
Engineering Division, of the U.S. Department of Energy under
Contract No. DE-AC02-05CH11231. RC acknowledges support from EPSRC
(U.K.) through Grant No. EP/H014934/1.

%\clearpage

%%%%%   APPENDIX   %%%%%%

%%%%%   APPENDIX   %%%%%%

\section*{appendices}
\appendix

\subsection{Parent orthorhombic setting for
$\alpha,\beta,\gamma$-Li$_2$IrO$_3$}
\label{sec:vectors}
In this section, we define simple idealizations of the Ir lattices in the
crystals, by taking oxygen octahedra to have ideal cubic symmetry. This
provides a pedagogically clearer description of the 3D lattices. For the
layered $\alpha$-Li$_2$IrO$_3$ monoclinic structure, our definition of parent
orthorhombic axes is a key step in our prediction of its magnetic order, as
discussed in the text.

We use a coordinate system based on the parent orthorhombic axes shown in Fig.
1. These vectors, which are the conventional crystallographic axes for
$\beta,\gamma$-Li$_2$IrO$_3$, are related to the Ir-O $x,y,z$ axes by
\begin{equation}
\bm{a} = (2,2,0) , \quad \bm{b} = (0,0,4), \quad \bm{c} = (6,-6,0).
\end{equation}
In the equation above we have written the $a,b,c$ vectors in terms of the
Cartesian (cubic orthonormal) $x,y,z$ coordinate system. The
$\hat{x},\hat{y},\hat{z}$ lattice vectors in this coordinate system are
defined as the vectors from an iridium atom to its neighboring oxygen atoms in
the idealized cubic limit, with the unit of length being the Ir-O distance.
Nearest neighbor bonds in the resulting Ir lattice have length $\sqrt{2}$, and
second neighbors are at distance $\sqrt{6}$.

For each lattice, we express its Bravais lattice vectors, as well as each of
its sites of its unit cell, in terms of the $a,b,c$ axes. A given vector or
site, written as $(n_a,n_b,n_c)$, is converted to the Cartesian coordinate
system by $(n_x,n_y,n_z) = n_a \bm{a} + n_b \bm{b}+ n_c \bm{c}$. The
conventional unit cell in the orthorhombic setting, which contains 16 sites,
is found by combining the primitive unit cell with the Bravais lattice
vectors.

\textbf{$\beta$-\li213 hyperhoneycomb} lattice ($n{=}0$ harmonic honeycomb),
space group $Fddd$ (\#70):

Primitive unit cell (4 sites):
\begin{align}
\bigg(0, 0, 0\bigg); \ \left(0, 0, \frac{1}{6}\right); \ \left(\frac{1}{4},
\frac{-1}{4}, \frac{1}{4}\right); \ \left(\frac{1}{4}, \frac{-1}{4},
\frac{5}{12}\right)
\end{align}

Bravais lattice vectors (face centered orthorhombic):
\begin{equation} \left(\frac{1}{2}, \frac{1}{2}, 0 \right); \
\left(\frac{1}{2}, -\frac{1}{2},0 \right); \ \left(\frac{1}{2}, 0, \frac{1}{2}
\right) . \end{equation}

\textbf{$\gamma$-\li213 stripyhoneycomb} lattice ($n{=}1$ harmonic honeycomb),
space group $Cccm$ (\#66):

Primitive unit cell (8 sites):
\begin{align}
&\bigg(0, 0, 0\bigg); \ \left(0, 0, \frac{1}{6}\right); \ \left(\frac{1}{4},
\frac{-1}{4}, \frac{1}{4}\right); \ \left(\frac{1}{4}, \frac{-1}{4},
\frac{5}{12}\right); \nonumber\\
&\left(0, 0, \frac{1}{2}\right); \ \left(0, 0, \frac{2}{3}\right); \
\left(\frac{1}{4}, \frac{1}{4}, \frac{3}{4}\right); \ \left(\frac{1}{4},
\frac{1}{4}, \frac{11}{12}\right)
\end{align}

Bravais lattice vectors (base centered orthorhombic):
\begin{equation} \left(\frac{1}{2}, \frac{1}{2}, 0 \right); \
\left(\frac{1}{2}, -\frac{1}{2},0 \right); \ \bigg( 0, 0, 1\bigg) .
\end{equation}

\textbf{$\alpha$-\li213 layered honeycomb} lattice ($n{=}\infty$ harmonic
honeycomb), space group $C2{/}m$ (\#12):

To discuss the layered honeycomb $\alpha$-\li213 polytype within the context
of its 3D cousins, we must first set up a single global coordinate system. The
two 3D lattices are captured, up to minute distortions, by the same parent
simple-orthorhombic coordinate system of $a,b,c$ axes.

The $\alpha$ polytype however has monoclinic symmetry and is conventionally
described by a set of monoclinic axes, which we denote
$\bm{a_m},\bm{b_m},\bm{c_m}$. The parent orthorhombic $\bm{a},\bm{b},\bm{c}$
axes defined above are distinct from the conventional monoclinic axes used to
describe this $C2{/}m$ crystal. Here we define an orthorhombic coordinate
system from a higher-symmetry idealization of these monoclinic axes,
by taking $\bm{a_o}=\bm{a_m}+\bm{c_m}, \ \bm{b_o}=\bm{a_m}-\bm{c_m}, \
\bm{c_o}=\bm{2b_m}$. The $a_o,b_o,c_o$ notation here signifies that, up to the
distortions of oxygen octahedra, the resulting $a,b,c$ axes are identical to
the orthorhombic axes of the $\beta$ and $\gamma$ polytypes. This
higher-symmetry idealization consists of the approximation that $|a_m|=|c_m|$,
which is wrong in the physical lattice\cite{OMalley2008} only by about 1\%.
The transformation between the conventional monoclinic axes and the universal
orthorhombic axes is also described by the coordinate notation as
\begin{equation} \bm{a_m}= \left(\frac{1}{2}, \frac{1}{2}, 0 \right); \
\bm{b_m}=\left(0,0,\frac{1}{2}\right); \ \bm{c_m}=\left(\frac{1}{2},
-\frac{1}{2}, 0 \right) . \end{equation}
The $a,b,c$ coordinate system preserves the key features used to discuss the
other lattices, namely that bonds lying along the $c$ axis carry Kitaev
coupling $b=z$, while remaining bonds are diagonal to the $a,b,c$ axes and
form the $d$-bonds zigzag chains. Equivalently, we choose a right handed
orthorhombic coordinate system, with the $c$ axis as the unique axis along
which one third of Ir-Ir bonds are aligned, and the $b$ axis as the unique
axis along which one third of Ir-O bonds are aligned.

Primitive unit cell (2 sites, denoted $A$ and $B$):
\begin{align}
\bigg(0, 0, 0\bigg);\ \left(\frac{1}{4}, -\frac{1}{4}, \frac{1}{12}\right)
\end{align}

Bravais lattice vectors, here denoted as $a_1,a_2,a_3$:
\begin{equation} a_1= \left(\frac{1}{2}, -\frac{1}{2}, 0 \right); \
a_2=\left(-\frac{1}{4}, \frac{1}{4},\frac{1}{4} \right); \
a_3=\left(\frac{1}{2}, \frac{1}{2}, 0 \right) \end{equation}
where the first two vectors span the 2D honeycomb plane.
These vectors are all of the same length ($\sqrt{6}$ in units of Ir-O
distance), and span the six second neighbors within a honeycomb plane, plus
one of the two additional pairs of sites on adjacent planes which are at the
same distance, given by vectors $\pm a_3 = \pm(\hat{x}+\hat{y}+2\hat{z})$ (the
remaining pair belongs to the opposite sublattice).

Within a honeycomb plane, the nearest neighbor vectors from $A$ to $B$ are
$r_1,r_2,r_3$, with $r_3={-}r_1{-}r_2$ and
\begin{equation} r_1 =\left(-\frac{1}{4}, \frac{1}{4}, \frac{1}{12}\right); \
r_2 =\left(\frac{1}{4}, -\frac{1}{4}, \frac{1}{12}\right). \end{equation}
The Bravais vectors above are related by $a_1=r_2-r_1$, $a_2=r_1-r_3$. For
reference we also note these Ir-Ir vectors in the Ir-O coordinate system, $r_1
= -\hat{y}+\hat{z}$, $r_2=\hat{x}-\hat{z}$, $r_3 = -\hat{x} +\hat{y}$. This
immediately implies that the Kitaev labels for $(r_1,r_2,r_3)$ bonds are
$(x,y,z)$ respectively.

\textbf{Zigzag chain} as basic structural unit:

The 1D zigzag chain is composed of sites $A$ and $B$,
\begin{align}
\bigg(0, 0, 0\bigg);\ \left(\frac{1}{4}, -\frac{1}{4}, \frac{1}{12}\right) ,
\end{align}
together with a single (1D) Bravais lattice vector,
\begin{equation} a_1= \left(\frac{1}{2}, -\frac{1}{2}, 0 \right).
\end{equation}
The reflection $b\rightarrow -b$ takes this zigzag chain to its
symmetry-equivalent partner, in which the minus sign in the two equations
above is replaced by a plus sign.

In this notation it is evident that the zigzag chains forms the basic
structural unit in all three \li213 polytypes. In each lattice, sites are
naturally partitioned into pairs which match this zigzag chain unit cell, and
each lattice contains the chain's Bravais lattice vector. The magnetic
Hamiltonian on each lattice is constructed as the sum of zigzag chain
Hamiltonians plus inter-chain interaction terms.

\subsection{Ising interactions} 
\label{sec:appendixising}
\begin{figure}[]
\includegraphics[width=0.91\columnwidth]{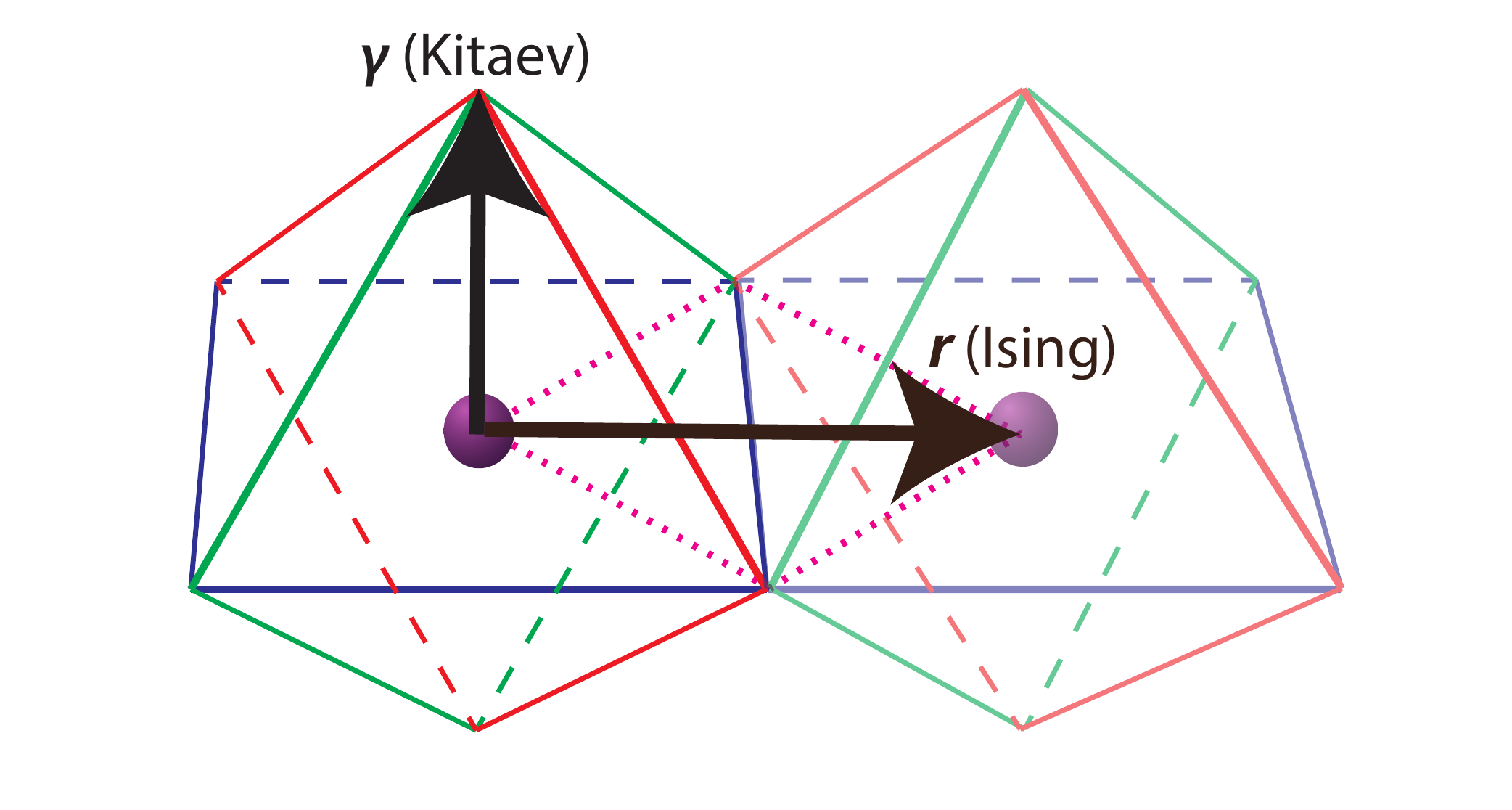}
\caption[]{{\bf Visualization of geometry of Kitaev and Ising exchanges.}
The two neighboring Ir sites (purple spheres), with surrounding oxygens
(vertices of octahedra), are shown. The oxygen octahedra of neighboring Ir
sites are edge-sharing in these structures. The axes for the anisotropic
interaction terms are then determined as follows (see the discussion following
Eq. 1 of the main text for details). The Ising interaction axis $\vec{r}$ is
the vector connecting the two Ir sites. The Kitaev interaction axis $\gamma$
is perpendicular to the plane which contains $r$ and the shared octahedra
edge. For both interaction terms, the coupling axis for the quadratic spin
interaction is defined as an axis with no orientation; here it is shown as an
arrow (with an arbitrary direction of the arrow head) for ease of
visualization.
}
\label{fig:octahedra}
\end{figure}

The Ising term defined in Eq.~1 is distinct from any combination of Kitaev and
Heisenberg exchanges. (The geometry is visualized in
Fig.~\ref{fig:octahedra}.) It can be related to the ``off-diagonal'' symmetric
interactions which have recently appeared in the
literature\cite{Kee2014,Imada2014,vandenBrink2014} under the symbols $\Gamma$
or $D$. For instance, if on a $z$-bond one writes the term ${+}\Gamma (S_i^x
S_j^y + S_i^y S_j^x)$, then the triplet $KJI$ reproduces $JK\Gamma$ by setting
$(K,J,I) = (K{-}\Gamma,J{+}\Gamma,{-}2\Gamma)$.
The bond-Ising interaction may be preferred as its definition, unlike
$\Gamma$,
 is independent of coordinate system.

\begin{figure}[]
\includegraphics[width=0.91\columnwidth]{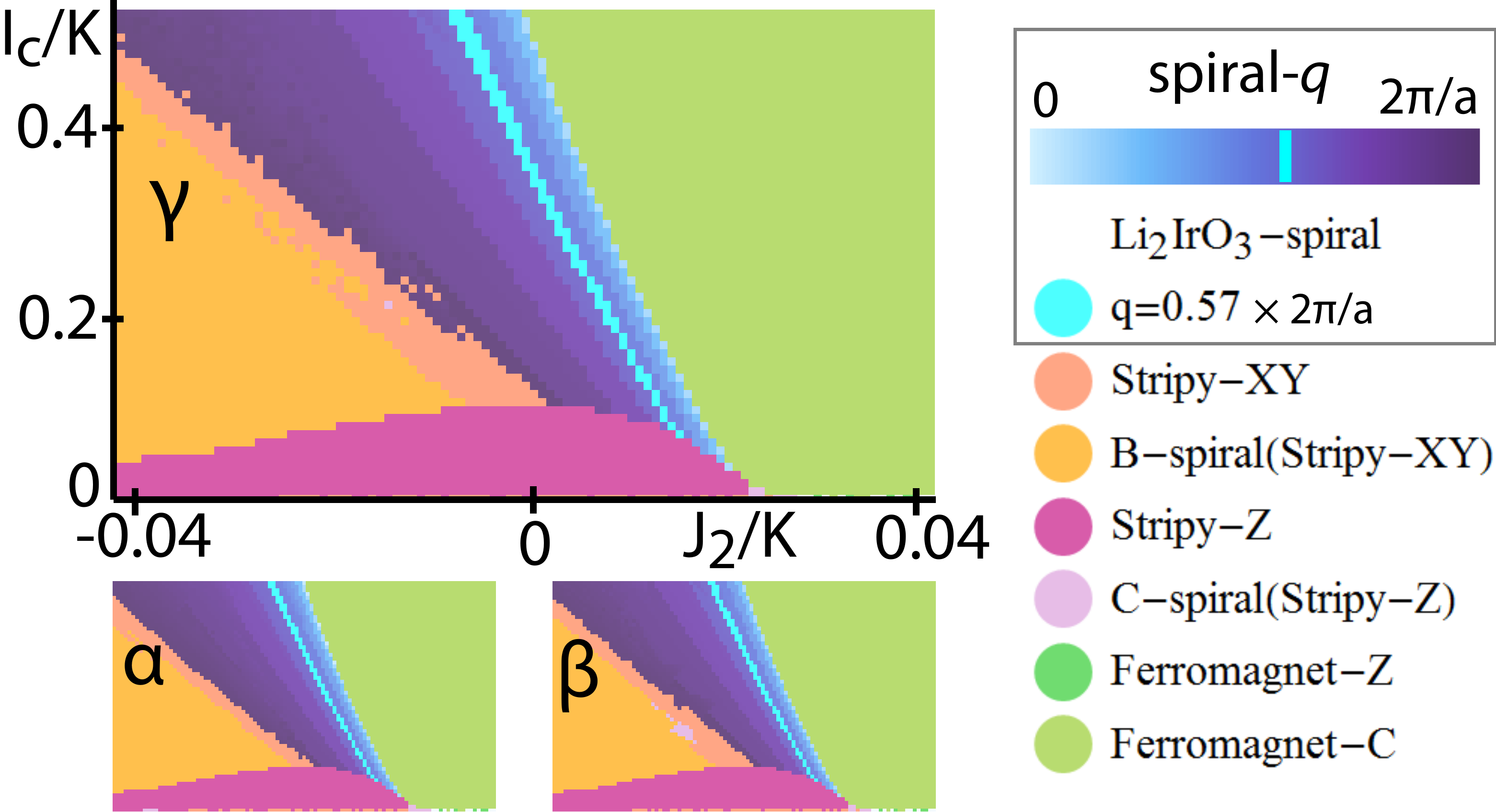}
\caption[]{%{\bf Additional phase diagram.}
Here we observe that for $J=|K|/20$, finite $I_c<0$ is required regardless of
the sign or magnitude of $J_2$.
}
\label{fig:phases4}
\end{figure}

In Eq.~1 we have included the Ising coupling only on $c$-bonds, for the
following reasons.
First consider the coplanar spiral mode. Since $r_{i j} \perp \gamma_{i j}$
and on $d$-bonds $\gamma_{i j}=(\hat{x},\hat{y})$, the $d$-bond $r_{i j}$ take
values $((\hat{y},\hat{x}){\pm} \hat{z})/\sqrt{2}$, projecting $I_d$ into a
Heisenberg-Kitaev term when $\langle S^z \rangle{=}0$.
In contrast $I_c$ couples spin component
$\hat{c}{=}(\hat{x}{-}\hat{y})/\sqrt{2}$ and helps stabilize the spiral
(Appendix Fig.~\ref{fig:phases4}). Second, we observe that the
experimentally-observed pattern of non-coplanar tilts is not favored by the
$d$-bonds Ising exchange, whose $r_{i j}$ orientations favor a different
symmetry breaking pattern. The correct $S^b$ tilts are instead stabilized by
the $K_c$ Kitaev term.

\subsection{Details of relation between Ising and $\Gamma$ terms}
\label{app:gamma}
We show more explicitly how the off-diagonal symmetric interaction term,
sometimes called the ``$\Gamma$'' exchange, can be made equivalent to the
Ising term introduced above by appropriately modifying the strength of the
Kitaev and Heisenberg couplings. This can be seen by writing the spin
interaction matrix $J^{a,b}$ for the interaction $S^a J^{a,b} S^b$ (summation
implied) of neighboring spins. Let us again write it in the $KJI$ and
$JK\Gamma$ notations for the interaction on a $c$-bond, in the $x,y,z$ basis,
\begin{equation}
\left( \begin{array}{ccc}
\frac{1}{2}I_c{+}J & -\frac{1}{2}I_c & 0 \\
-\frac{1}{2}I_c & \frac{1}{2}I_c{+}J & 0 \\
0 & 0 & K{+}J \end{array} \right)
\longleftrightarrow
\left( \begin{array}{ccc}
J & \Gamma_c & 0 \\
\Gamma_c & J & 0 \\
0 & 0 & K{+}J \end{array} \right)
\end{equation}
where we have kept the $c$ subscript on $I_c$ and $\Gamma_c$ to denote that
these are the parameters for the $c$-type bond.
The set of interaction matrices spanned by $K,J,I$ is equivalent to that
spanned by $J,K,{\pm}\Gamma$. In particular, our $K,J,I_c$ model, with Ising
interactions on $c$-bonds, is related to a $K,J,\Gamma_c$ model with
off-diagonal $\Gamma_c$ couplings on $c$ bonds.

The bond-Ising interaction may be preferred for two reasons. First, its
geometric definition, coupling the spin component along the Ir-Ir bond, is
independent of coordinate system and thus free of sign ambiguities. In
contrast, distinguishing ${+}\Gamma$ from ${-}\Gamma$ is coordinate-dependent.
This is most evident for the $x$ and $y$ bonds on the 3D lattices, where in
the $\Gamma$ notation the interaction appears with a positive sign on half of
the $x$-bonds and a negative sign on the remaining $x$-bonds. In contrast, the
Ising term directly sets the coupled spin component to the direction of the
displacement vector between the two sites, and is invariant to the vector's
sign. Second, the Ising coupling, of spin components along the bond,
transparently indicates
that this exchange is symmetry-permitted even for ideal O$_6$ octahedra.

\subsection{Details of the semiclassical solution}
Here we give technical details for the semiclassical solution. First note that
the 16-site unit cell of the orthorhombic axes contains 4 sites along the
spiral propagation direction $a$; in contrast, the zigzag-chain 1D Bravais
vector $a_1$ spans two sites. Hence a wavevector in units of $\pi/a_1$ is
roughly analogous to one in units of $2\pi/a$.

For all three lattices, we use an 8-site unit cell with a base-centered
orthorhombic Bravais lattice. In this choice of unit cell, the Brillouin zone
is rotated (by 45 degrees) and doubled in area from the BZ associated with the
conventional orthorhombic coordinate system; e.g.\ it extends from $-2\pi/a$
to $+2\pi/a$ along the $a$-axis. We perform numerical minimization by defining
a $\pi/8$-spaced grid in
the Brillouin zone and then using the constrained
minimization algorithm of
Broyden-Fletcher-Goldfarb-Shanno\cite{Broyden1970,BFGS},
independently starting at each grid point.

Let us write the explicit process of solution for the wavevector within the
Fourier transform (FT). For concreteness we focus on the minimal parameters
$(K,J,I_c)=(-12,0.6,-4.5)$ meV, on the $\beta$ (hyperhoneycomb) lattice. This
Hamiltonian is minimized at $\vec{q}=0.57 \times 2\pi/|a| \times \hat{a}$. The
FT ground state at this wavevector, energy -14.8 meV, has ordered spin moment
$\vec{S}\propto \hat{c}\pm i 0.48 \hat{a}$, where the $\pm$ sign alternates
between successive sites in the unit cells (shown above) when they are listed
in order of their $c$ coordinate. The second excited state at this wavevector,
energy -12.1 meV is capable of mixing with this ground state, and exhibits a
wavefunction $\pm \hat{b}$ where this distinct $\pm$ symbol is chosen to give
the same sign on two sites connected by a $c$-bond, and opposite sign on two
sites connected by a $d$-bond; in other words, it alternates in pairs when
sites are listed by their $c$ coordinate. Observe that these definitions of
sign structure are consistent with the definition of the wavefunction given in
the text, Eq.~\ref{eq:spinfn}.

The mixing mode energy can be tuned towards the ground state, for example in
the nearby set of parameters with bond-strength anisotropy in the Kitaev term,
$(K_c,K_d,J,I_c)=(-13.2,-11,0.6,-4.5)$ (in meV), the ground state coplanar
mode has energy -13.8 meV, and the tilt mode is its first excited state, at
energy -13.5 meV higher. This combined noncoplanar state is found on all three
lattices. As discussed in the text, it agrees with the spiral order observed
experimentally on both the $\beta$ and the $\gamma$ polytypes.

Finally, we note that in labeling the phases within the numerically-computed
phase diagrams, we have used features which are invariant across the phase,
such as the ordering wavevector and the pattern across lattice sites. Due to
the strong spin orbit coupling which microscopically generates the model
Hamiltonian, and the associated Hamiltonian-level breaking by the Kitaev as
well as the Ising terms of any spin symmetries, the spin moment ordering
direction on the Bloch sphere is not a robust measure of a phase. In
particular, this Bloch sphere direction of the ordered spin moment generally
varies smoothly as parameters are varied, within a given collinear
antiferromagnetic or ferromagnetic phase.

\subsection{Details of the 1D zigzag-chain solution}
\label{app:chain}
Here we present the full solution of the zigzag-chain model within the
$x,y$-coplanar ansatz shown in the text.
The quickest route to deriving the energy function Eq.~\ref{eq:Eminus} is to
plug in the spin-spin correlations into the Hamiltonian Eq.~\ref{eq:chain}.
The nearest-neighbor correlations are given in Eq.~\ref{eq:spincorrelations};
the second neighbor correlations are $\langle \vec{S}_r{\cdot}\vec{S}_{r+a_1} \rangle = \cos(q
a_1)$. These two equations are sufficient for solving the model.

Alternatively, plugging in the ansatz Eq.~\ref{eq:spinansatz} into the
Hamiltonian Eq.~\ref{eq:chain} gives the following energy function,
\begin{align}
E_{\text{1D}} = \sum_{r{=}n a_1} & \bigg[\
 J_2 \,   \big(2\cos(\theta)  \big) \notag\\
& + K  \, \bigg(\ \cos\left(\theta/2\right) \cos(f_-(r)) \notag\\
&\qquad \ + \sin\left(\theta/2\right) \sin(f_+(r))  \bigg) \notag\\
&+ J\  \, \big(2 \cos\left(\theta/2\right) \cos(f_-(r)) \big)\phantom{\big[}
 \bigg]
 \notag\\
f_{\pm}(r) &=\big[ (\phi_A \pm \phi_B) + r(q_A \pm q_B) \big]
\end{align}
with $\theta = a_1 q_B$.
Performing the average over 1D Bravais lattice sites $r=n a_1$, we observe
four possibilities. If $ q_A= q_B\neq 0,\pi/a_1$, then the term with $f_+$
vanishes, while $f_-$ are replaced by $(\phi_A-\phi_B)$. This co-rotating
spiral is set by the interplay of primarily Heisenberg first and second
neighbor exchanges, requires the typical geometrical frustration here encoded
by $J$ and $J_2$ of the same sign, and is the typical spiral one expects from
frustrated Heisenberg models. If $ q_A=- q_B\neq 0,\pi/a_1$, then the terms
with $f_-$ vanish, while $f_+$ are replaced by $(\phi_A+\phi_B)$. This is the
counter-rotating spiral.
The final possibilities are $\theta=\pm \pi$, leading to the stripy
antiferromagnet, or $\theta=0$, leading to the ferromagnet (in both cases
$f_\pm$ are replaced by $(\phi_A\pm\phi_B)$), discussed above.

When studying the counter-rotating spiral, it is important to keep in mind the
behavior of the phases under lattice translations.
Due to the counter-rotation, here the \textit{average} phase is the physical
quantity; the arbitrary ``overall phase'' of the spiral, freely modified (for
incommensurate $q$) by shifting $r$, is then the \textit{difference} of phases
$\phi_A{-}\phi_B$. We may choose the phases $\phi_A{=}\phi_B{=}\pi/4$ to
satisfy $\phi_A{+}\phi_B{=}\pi/2$, keeping in mind that shifting the overall
phase does not permit these phases to simultaneously be set to zero.

The stabilization of the spiral by Kitaev interactions can also be observed
via Eq.~\ref{eq:spincorrelations} by fixing $\phi_A{+}\phi_B=\pi/2$. While the
Heisenberg correlator vanishes, the spin component matching the Kitaev bond
type exhibits nonzero correlations, $\langle S^x_r S^x_{r+r_1} \rangle_x
=(1/2)\sin(\theta/2)$.

%\clearpage
%%%%%   BIBLIOGRAPHY   %%%%%%
\bibliography{IridatesCitations}

\end{document}